\documentclass[useAMS,usenatbib,usegraphicx]{mn2e}

\def\beq{\begin{equation}}
\def\eeq{\end{equation}}
\def\Omm{{\Omega_{\rm m}}}

\def\Oml{{\Omega_{\Lambda}}}
\def\tE{\theta_{\rm E}}
\def\rE{r_{\rm E}}
\def\c2{c_{200}}
\def\cv{c_{\rm vir}}
\def\r2{r_{200}}
\def\rv{r_{\rm vir}}
\def\rs{r_{\rm s}}
\def\Mv{M_{\rm vir}}
\def\M2{M_{200}}
\def\zs{z_{\rm sat}}
\def\fs{f_{\rm sat}}
\title[Tidal Enhancement of Cluster Concentrations]{Concentrating the Dark
Matter in Galaxy Clusters through Tidal Stripping of
Baryonically-Compressed Galactic Halos} \author[Rennan Barkana and
Abraham Loeb]{Rennan Barkana$^{1,2}$ and Abraham
Loeb$^{3}$\thanks{E-mail: barkana@wise.tau.ac.il (RB);
aloeb@cfa.harvard.edu (AL)}\\ $^{1}$ Division of Physics, Mathematics
and Astronomy, California Institute of Technology, Mail Code 130-33,
Pasadena, CA 91125, USA \\ $^{2}$ Raymond and Beverly Sackler School
of Physics and Astronomy, Tel Aviv University, Tel Aviv 69978,
Israel\\ $^{3}$Astronomy Department, Harvard University, 60 Garden
Street, Cambridge, MA 02138, USA}

\begin{document}

\pagerange{\pageref{firstpage}--\pageref{lastpage}} \pubyear{2008}
\maketitle
\label{firstpage}
\begin{abstract}
Gravitational lensing observations of massive X-ray clusters imply a
steep characteristic density profile marked by a central concentration
of dark matter. The observed mass fraction within a projected radius
of 150 kpc is twice that found in state-of-the-art dark matter
simulations of the standard $\Lambda$CDM cosmology. A central baryon
enhancement that could explain this discrepancy is not observed,
leaving a major puzzle. We propose a solution based on the merger
histories of clusters. A significant fraction of the final dark matter
content of a cluster halo originates within galaxy-sized halos, in
which gas can cool and compress the dark matter core to high
densities. The subsequent tidal stripping of this compressed dark
matter occurs in denser regions that are closer to the center of the
cluster halo.  Eventually, the originally cooled gas must be dispersed
into the intracluster medium through feedback, for consistency with
observations that do not find central baryon enhancements in
clusters. Still, the early adiabatic compression of the galactic dark
matter leaves a net effect on the cluster. Using a simple model for
this process, we show that the central cluster profile is
substantially modified, potentially explaining the observed
discrepancy.
\end{abstract}


\section{Introduction}\label{intro}

Recent observations have confirmed our basic understanding of
cosmology and showed an impressive consistency with the predictions of
the standard $\Lambda$CDM model \citep[e.g.,][]{SNV06,BAO07,wmap}. In
this model, a cosmological constant dominates the cosmic mass budget
today, but galaxies and other structures were assembled earlier,
primarily out of cold dark matter. This medium of non-interacting, low
velocity-dispersion particles, started out with small Gaussian density
perturbations that were subsequently enhanced by gravity. While the
model successfully matches observations of the large-scale
anisotropies of the cosmic microwave background and the large-scale
structure in galaxy surveys, it is also important to test its validity
on smaller scales. The abundance and structure of non-linear objects
are potentially sensitive probes of the properties of dark matter
(e.g., whether it is cold) and of the density fluctuations (e.g.,
whether they are Gaussian). However, gas cooling and astrophysical
feedback complicate the interpretation of observations regarding the
dark matter distribution in galaxies. Thus, it is most attractive to
study the largest virialized objects, namely X-ray clusters, in which
most of the gas is too hot and rarefied to cool and is thus expected
to trace the gravitational potential.

The mass profiles of galaxy clusters can be measured directly through
gravitational lensing. Observations of the most massive clusters now find
dozens of multiply-imaged background sources, allowing a precise
measurement of the central 2-D mass distribution in each cluster as
projected on the sky. Also crucial for characterizing each cluster is its
total virial mass, which can be measured precisely by supplementing the
central strong lensing signal with weak lensing distortions measured out to
the cluster edge. It is useful to characterize the total, projected profile
with one scale, the effective Einstein radius $\rE$ (or angle $\tE$)
defined so that a circle of that radius around the cluster center contains
a mean enclosed surface mass density $\bar{\Sigma}$ equal to the critical
density for lensing, $\Sigma_{\rm cr}=[c^2/(4\pi G)]D_{\rm OS}/(D_{\rm OL}
D_{\rm LS})$, where $D$ denotes various angular diameter distances
(Observer-Source, Observer-Lens, and Lens-Source). This definition is
motivated by the Einstein ring radius of an axisymmetric lens, but is
nonetheless a useful measure of the central matter content even for
asymmetric clusters.

Current N-body simulations of galaxy clusters in $\Lambda$CDM produce
samples of thousands of halos with virial mass $\Mv>10^{14} M_\odot$. These
simulations are becoming sufficiently large and detailed to yield the
predicted spread in cluster halo parameters, and to allow a quantitative
assessment of the inherent bias in observing clusters in projection and
selecting them by lensing cross-section \citep{Hennawi,Neto}. In general,
the density profiles of the simulated clusters are relatively shallow and
seemingly at odds with recent careful lensing studies of massive clusters
\citep{Kneib,Gavazzi,Br05b,kling,Limousin,Bradac,halkola08,umetsu}.

This discrepancy was recently highlighted and quantified by \citet{BB08},
who carefully compared observations of four well-constrained massive
clusters to the predictions of the numerical simulations. They emphasized
the importance of comparing directly the projected 2-D mass distributions
in the observations and the simulations, using the virial mass ($\Mv$) and
the effective Einstein radius ($\rE$) as two numbers that characterize the
degree of concentration in each cluster halo. They derived the theoretical
predictions for cluster lensing in $\Lambda$CDM by starting with the
distribution of 3-D halo profiles measured by \citet{Neto} in the
Millennium simulation, and then correcting it for lensing and projection
biases based on \citet{Hennawi}. Comparing the resulting distribution with
the observed $\rE$ for four clusters -- A1689, Cl0024, A1703, and RXJ1347
-- and including the expected spread in profiles as well as the measurement
errors, they found that each cluster was discrepant at the 2--$\sigma$
level (all with an unusually large $\rE$ given $\Mv$), yielding a combined
4--$\sigma$ discrepancy. \citet{Duffy} recently found that simulated
cluster concentrations are even lower when using the most updated
cosmological parameters (which have a lower power spectrum normalization
than assumed by \citet{Neto}), though the effect for the most massive
clusters is only at the level of $\sim 10\%$.

\citet{BB08} suggested that gas physics is unlikely to affect significantly
the Einstein radius of a cluster. This radius of $\sim 150$ kpc is
typically observed to enclose a projected mass of $\sim 2 \times
10^{14} M_\odot$, or a mass of $\sim 1 \times 10^{14} M_\odot$ within
the same 3-D radius. Using the simple model of adiabatic compression
\citep{blumenthal}, they estimated that the observed 3-D mass within
the Einstein radius can be obtained if gas cooling increases the
enclosed baryonic fraction within this radius to $\sim 1/3$, twice the
cosmic baryon fraction.  Indeed, hydrodynamic simulations produce
clusters that are as centrally concentrated as those observed, likely
due to their ``overcooling'' problem which produces just such an
increase in the central baryon fraction, with most of it in stars
\citep{Kravtsov, Rozo}. An increase of this sort apparently does not
occur in real clusters, where the baryonic (gas$+$stellar) mass
fraction within the Einstein radius is below the cosmic value
\citep[e.g.,][]{Lin, LaRoque, Doron, Afshordi, Vikhlinin}. Thus,
\citet{BB08} concluded that cluster halo profiles present perhaps the
clearest, most robust, current conflict between observations and the
standard $\Lambda$CDM model. Subsequent work has generally supported
this conclusion \citep[e.g.,][]{Oguri,OB09,Zitrin}, though only a
large unbiased cluster sample with precise strong and weak lensing
measurements would be completely conclusive.

In this paper we propose a novel process that could resolve the
apparent discrepancy between cluster observations and existing
$\Lambda$CDM simulations. A significant fraction of the final dark
matter content of a cluster halo originates within galaxy-sized halos,
in which gas can cool and compress the dark matter core to high
densities. In \S~2 we develop a simple model for this adiabatic
compression and for the subsequent tidal stripping of the dark matter
within the cluster halo. We then show in \S~3 that the central mass
profile of the galaxy cluster is substantially modified by compression
of the galactic halos that it swallows, even if the cooled galactic
baryons are later redistributed throughout the cluster. We show
quantitatively that this can potentially explain the observed
discrepancy. Finally, we summarize our conclusions and caveats in
\S~4.

We note that \citet{Moore} found in a simulation of a Galactic halo
that cooling produced a more highly concentrated dark matter profile
for the host halo, but there, as in the cluster simulations, this may
have been due directly to adiabatic compression in response to the
large concentration of baryons in the center of the host
halo. \citet{Dolag} included radiative cooling and stellar feedback in
simulations of cluster formation, and while they focused on the
structure of the galactic subhalos, they also found a $\sim 15\%$
increase in the effective concentration of the cluster halo; this,
however, may still be partly due to an increased baryon content near
the cluster center. Also, \citet{SL} artificially eliminated the
over-cooling problem in their cluster simulations and found only a
very minor effect of the baryons on the total mass profile in this
case; however, their artificial scheme may have also eliminated the
effect we analyze. Our analytical approach allows us to cleanly
separate the effect of density-enhanced tidal stripping from a simple
overall adiabatic compression of the cluster halo.

\section{Model}

\label{S:model}

We assume the standard $\Lambda$CDM cosmology \citep{wmap}, with a
dimensionless Hubble parameter $h=0.7$ and density parameters
$\Omega_m=0.28$ (dark matter plus baryons), $\Omega_\Lambda=0.72$
(cosmological constant), and $\Omega_b=0.046$ (baryons). We also denote the
cosmic baryon fraction by $f_b \equiv (\Omega_b/\Omega_m) = 0.16$.

Consider a halo that virialized at redshift $z$ in a flat $\Lambda$CDM
universe. The critical density at $z$ is \beq \rho_{\rm c}^z = \frac{3
H_0^2} {8 \pi G} \left[\Omm (1+z)^3 + \Oml \right] \ . \eeq Numerical
simulations of hierarchical halo formation indicate a roughly
universal spherically-averaged density profile for virialized halos
\citep[hereafter NFW]{NFW}: \beq \rho(r)=\rho_{\rm c}^z\,
\frac{\delta_c} {\frac{r}{\rs} \left(1+ \frac{r}{\rs} \right)^2}\ ,
\eeq where the radius $r$ is divided by the scale radius $r_s =
\rv/\cv$ with $\rv$ being the virial radius, and the characteristic
density $\delta_c$ is related to the concentration parameter $\cv$ by
\beq \delta_c= \frac{\Delta_c}{3} \frac{\cv^3}
{\ln(1+\cv)-\cv/(1+\cv)} \ , \eeq where $\Delta_c$ is the virial
density in units of $\rho_{\rm c}^z$. For a halo of virial mass $\Mv$
at a given redshift $z$, the profile is fully specified by the
parameters $\Delta_c$ and $\cv$. We adopt the convention of a fixed
$\Delta_c = 200$ at all redshifts, for consistency with the simulation
analyses whose results we use.

We adopt a simple analytical model of tidal stripping that, in
particular, has been previously used to understand how the density
profiles of satellite sub-halos produce the central NFW profile of
their final host halo \citep{Syer}. In this model, material originally
at a radius $\xi$ within the satellite ends up, after tidal stripping,
at a radius $r$ within the host halo so that the mean enclosed
satellite density within $\xi$ equals the mean enclosed host density
(before the stripping) within $r$:
\beq
\bar{\rho}_{\rm host}^0(r) =
\bar{\rho}_{\rm sat}(\xi) \ . \label{e:noac}
\eeq
This condition corresponds to a resonance in dynamical frequencies
between the circular orbit of a mass element at $\xi$ around the
satellite and that of a circular satellite's orbit around the host
halo, inducing an energy transfer that strips the mass element from
the satellite. Setting the host tidal force equal to the internal
gravitational force within the satellite yields a very similar
stripping radius.

We combine this stripping model with the simple model of adiabatic
compression \citep{blumenthal} in which conservation of angular momentum
implies that the quantity $r M(r)$ (assuming spherical symmetry) is
fixed. We assume that both the host and each stripped satellite start out
with NFW profiles. In satellites within the mass range of galaxies, the
baryons cool and condense to the center, inducing a change in the
surrounding dark matter halo. Specifically, adiabatic compression moves a
mass shell initially at $\xi_{\rm i}$, containing a mass $M_{\rm
sat}(\xi_{\rm i})$, to a final radius $\xi_{\rm f} = \xi_{\rm i} M_{\rm
sat}(\xi_{\rm i}) / M_{\rm f}$, where the final enclosed mass is larger by
a factor 
\beq 
\frac{M_{\rm f}} {M_{\rm sat}(\xi_{\rm i})} = 1-f_b + f_b
\frac{M_{\rm sat}} {M_{\rm sat}(\xi_{\rm i})}\ , \label{eq:Mf}
\eeq 
where $M_{\rm sat}$ without an argument denotes the total virial mass of
the satellite. The stripping model then implies that the same mass shell
ends up at a radius $r_{\rm f}$ in the host cluster halo, where \beq
\bar{\rho}_{\rm host}^0(r_{\rm f}) = \bar{\rho}_{\rm sat}(\xi_{\rm i}) 
\times \left[ \frac{M_{\rm f}} {M_{\rm sat}(\xi_{\rm i})} \right]^4\ ,
\label{e:strip} \eeq in terms of the initial enclosed density
$\bar{\rho}_{\rm sat}(\xi_{\rm i})$ in the satellite (i.e., before the
adiabatic compression). The power of four on the right-hand side
results from the increased mass (one power) and decreased radius
(hence three powers in the density). Note that the cooled baryonic
cores of the galaxies are much denser than their surrounding dark
matter halos and so we have assumed that the baryonic cores are not
tidally stripped.

In deriving equation~(\ref{eq:Mf}) we have assumed that in galactic
halos, where the cooling time of the virialized gas is much shorter
than the Hubble time, the full baryonic content of the halos initially
cools toward the center of the halo, condensing the surrounding dark
matter before it gets stripped. Observationally, even the fraction of
galactic halo baryons that are in stars today is not well known, since
the hot baryons in halos are difficult to detect, while total masses
of galactic halos are difficult to measure accurately (and can be used
to obtain the total gas mass only with the added assumption of a halo
baryon fraction that equals the cosmic mean). The best estimate for
the total mass of stars and stellar remnants today, as a fraction of
the total baryonic mass that lies within virialized regions of
galaxies, is $\sim 10\%$, with an uncertainty of order $50\%$
\citep{Fuk}. For our own Milky Way galaxy, the disk and bulge may make
up as much as $40\%$ of the halo baryons
\citep{Xue}. Regardless of the precise fraction today, it is plausible
to assume that most of the gas in galaxies initially cooled and was
later expelled over time from the central region back into the halo,
through supernova or quasar feedback. Thus, we expect that if galactic
halos were stripped within the cluster relatively early, then star
formation and feedback did not have much time to operate prior to the
stripping. This scenario does not conflict with the fact that clusters
only virialized relatively recently, since we are focusing here on the
stripping that formed the inner regions of clusters, within a tenth of
the virial radius, and this likely occurred long before the entire
cluster virialized.

In clusters, the originally cooled gas must eventually have gotten
dispersed into the intracluster medium, since observations find only a
small fraction of cluster baryons residing near the center. In the
cluster environment, this gas redistribution can be facilitated by
interactions among galaxies or with the intracluster medium, in
addition to internal galactic feedback. For simplicity we assume that
the final baryon distribution is similar to that of the dark matter,
i.e., the final baryon fraction is uniform and equal to the cosmic
value.  Within our model above, this effectively means that the
satellites contribute only $M_{\rm sat}(\xi_{\rm i})$ to the mass
enclosed within the final cluster radius $r_{\rm f}$ (and not the full
$M_{\rm f}$).

In this picture, when the cooled galactic baryons get redistributed
throughout the final cluster halo, they may cause a partial adiabatic
expansion of the halo. However, this should roughly cancel the initial
adiabatic contraction of the surrounding halo material when the
satellites enter and are stripped; thus, we neglect both the initial
halo contraction and the later expansion, as we do not expect a
significant net effect. This is different from the main process that
we focus on, where the early adiabatic compression causes the
satellites' dark matter to be stripped at smaller cluster radii than
it would otherwise, leaving a net effect on the cluster in the end.

The stripping model that led to equation~(\ref{e:strip}) assumes that
the host halo is dominant, and that the satellites contribute only a
small fraction of the enclosed mass at the stripping radius. This
assumption breaks down, however, when the satellites dominate, and we
do expect such a regime; indeed, it is plausible (and indicated by
pure dark matter simulations) that the dense core of the cluster halo
arises entirely from the original cores of the accreted satellites,
since only lower-density material is accreted later onto the cluster
halo \citep{LP1,LP2}. In this high-density regime where the satellites
dominate relative to the pre-existing halo material, we expect the
satellite cores to simply settle in the host core, preserving in the
end their original densities, since they first adiabatically compress
and later re-expand. This limit is consistent with the equal-density
relation in equation~(\ref{e:noac}); while this equation was
originally derived for stripping onto a dominant host halo, we can use
it to capture the dominant satellite limit, if we obtain from it the
radius $r_{\rm f}$ at which the satellite mass ended up (reinterpreted
as contributing part of the halo profile rather than adding mass on
top of $\bar{\rho}_{\rm host}^0$). We leave for future work a detailed
analysis of the complex transition region between the two limits of
equation~(\ref{e:noac}) (density preservation) and
equation~(\ref{e:strip}) (density enhancement), and here we adopt a
simple interpolation between them. This is reasonable given our
limited goal of examining whether the halo profile can be
substantially modified at all. Thus, we determine the stripping radius
by solving (for a given $\xi_{\rm i}$): \beq \bar{\rho}_{\rm
host}^0(r_{\rm f}) = \bar{\rho}_{\rm sat} (\xi_{\rm i}) \times \left\{
f_{\rm sat}(r_{\rm f}) + [1-f_{\rm sat}(r_{\rm f})] \left[
\frac{M_{\rm f}} {M_{\rm sat}(\xi_{\rm i})} \right]^4 \right\} \, ,
\label{e:final} \eeq where $f_{\rm sat}(r_{\rm f})$ is the fractional
contribution of the satellites to the enclosed mass within the
stripping radius $r_{\rm f}$, and the right-hand side of this equation
interpolates between equation~(\ref{e:noac}) (valid in the limit
$f_{\rm sat}(r_{\rm f}) = 1$) and equation~(\ref{e:strip}) (valid when
$f_{\rm sat}(r_{\rm f}) = 0$).

In order to solve the model, we must determine various masses. First,
based on the host NFW profile in the absence of cooling and adiabatic
compression, we obtain the enclosed host mass at each radius, $M_{\rm
host}^0(r_{\rm f}) = \frac{4}{3} \pi r_{\rm f}^3 \bar{\rho}_{\rm
host}^0(r_{\rm f})$. We use equation~(\ref{e:noac}) to calculate the
fractional contribution of the satellites to this enclosed mass in the
absence of cooling, $f_{\rm sat}^0(r_{\rm f})$. Adiabatic compression
and stripping then replaces the satellite contribution by $M_{\rm f}$,
and thus in equation~(\ref{e:final}) we set
\beq f_{\rm sat}(r_{\rm f}) = \frac{M_{\rm f}} {[1-
f_{\rm sat}^0(r_{\rm f})] M_{\rm host}^0(r_{\rm f}) + M_{\rm f}}\ .
\label{e:fsat} \eeq After solving for $r_{\rm f}$, the final enclosed 
mass within this radius, after feedback redistributes the baryons, is
\beq
M_{\rm host}(r_{\rm f}) = [1- f_{\rm sat}^0(r_{\rm f})] M_{\rm
host}^0(r_{\rm f}) + M_{\rm sat}(\xi_{\rm i})\ . \label{e:fsatf}
\eeq

Within our simple model, the redistribution of mass depends only on
the density profile of the satellites, i.e., on the functional form of
$M_{\rm sat}(\xi_{\rm i}) / M_{\rm sat}$ versus $\bar{\rho}_{\rm
sat}(\xi_{\rm i})$, and not on the number of satellites or their
individual total masses. Since we are interested in the galactic
sub-halos that end up in the cluster, we can adopt the typical value
of accretion redshift $\zs$ and NFW concentration $\cv$ for such
halos, and effectively calculate stripping of one satellite that
contributes some total fraction $\fs$ of the host cluster mass. Within
the model, this single satellite represents the cumulative effect of
all the individual galactic satellites that merged into the final
cluster halo.

In order to quantify the effect of compressed galactic halos, we must
estimate the effective $\fs$, i.e., the fraction of cluster dark
matter that arrived from within galactic halos in which the baryons
were able to cool. We can obtain a theoretical estimate for the
fraction of cluster dark matter that passed in its merger history
through galactic halos, on the way to becoming part of the final
cluster halo. For concreteness, let us consider the progenitor
distribution at various redshifts of a $10^{15} M_\odot$ cluster halo
at $z \sim 0.2-0.4$. The extended Press-Schechter model \citep{bc91}
then implies that $\sim 25\%$ of the cluster mass was in halos with
masses in the range $10^{10}-10^{12} M_\odot$ at $z \sim 2.5$ (the
redshift that maximizes this fraction). A full merger tree would give
on average at least this value since additional cluster mass that was
outside this halo mass range at $z=2.5$ may have passed through
galactic halos at other redshifts.

We can also estimate the fraction $\fs$ from observations. A first
attempt might proceed as follows. Stars make up $\sim 1\%$ of the
total virialized mass of massive clusters \citep[e.g.,][]{Lagana08},
which correponds to $\sim 5\%$ of the total baryonic mass. To find the
baryon fraction that was associated with the galaxies in which these
stars formed, we must divide by their average star formation
efficiency. As noted above, the total stellar mass today is $\sim
10\%$ of the baryon mass within virialized regions of galaxies
\citep{Fuk}. This suggests that $\sim 50\%$ of the cluster gas was
processed through galaxies, and thus also a similar fraction of the
cluster's dark matter was contributed by stripped satellite galaxies
(assuming that the total baryon fraction of both these galaxies and
the final cluster is equal to the cosmic mean fraction).  However,
this estimate depends on the uncertain value of the star formation
efficiency. We can use metallicity measurements to obtain a more
direct estimate of $\fs$. The typical metallicity of the intracluster
medium in clusters at redshift $z\sim 0.3$ is 0.3 -- 0.4 of the solar
abundance \citep[e.g.,][]{Maughan}, while massive galaxies (in
clusters or the field) typically have a solar abundance or less
\citep[e.g.,][]{Ellison}, indicating that at least $30 - 40\%$ of the
cluster gas must have been processed in galaxies in order for the ICM
to reach its high metallicity value. Within our model, higher $\fs$
values lead to more highly concentrated cluster halos (see the next
section). Taking into account these various considerations, we
consider $\fs$ values in the range $20 - 40\%$.

We note that \citet{Oleg04} showed that the classic adiabatic
contraction model that we use tends to overestimate the effect of a
central baryon concentration on dark matter, compared to simulated
profiles. The overestimate in the mass profile, however, is under
$10\%$ at the radii that we focus on ($\sim 0.1 \rv$), thus justifying
our use of the simple model. The overestimate does increase at smaller
radii and is $\sim 50\%$ at $0.01 \rv$, implying that the profiles we
find below are less reliable in the innermost region.

\section{Results}

\label{S:res}

In this section we quantify the effect that adiabatic compression in
galactic satellites can have on the final density profile of the host
cluster halo. We compare our results to the four clusters considered
by \citet{BB08}, and also make use of their results for the
theoretical predictions. In particular, we adopt the NFW parameters
measured by \citet{Neto} for simulated halos, after correcting them
based on \citet{Hennawi} to obtain the effective parameters for the
population of lensing clusters, observed in projection; however, we
reduce $\cv$ by $10\%$ according to the recent results of
\citet{Duffy}. This yields a median $\cv=5.5$ for the most massive
clusters, with a 1--$\sigma$ range (for the effective projected $\cv$)
of 4--7.5 (approximately in a lognormal distribution). Studies based
on large numerical simulations \citep{Jing,Gao} have found for massive
halos a relatively weak decline of $\cv$ with increasing redshift, but
a more significant decline for galactic mass halos, with $\cv \sim 4$
for $\sim 10^{11}$--$10^{12} M_\odot$ halos at $z\sim 2$. In our
quantitative results, we adopt $\cv=4$ and $\zs=2$ for the satellites,
and consider hosts at $z=0.3$ with various concentration parameters.
As noted above, we assume that the dense core of the host arises
entirely from the satellite cores, so for each host $\cv$ this
normalizes the total satellite fraction $\fs$ of the cluster's virial
mass. In particular, we consider $\fs=20\%$ (which implies a host
$\cv=6.8$), $\fs=30\%$ ($\cv=7.5$), or $\fs=40\%$ ($\cv = 8.0$). These
host concentration parameters are somewhat high but still near or
within the expected 1--$\sigma$ range for cluster halos, as noted
above.

\begin{figure}%[th]
\includegraphics[width=84mm]{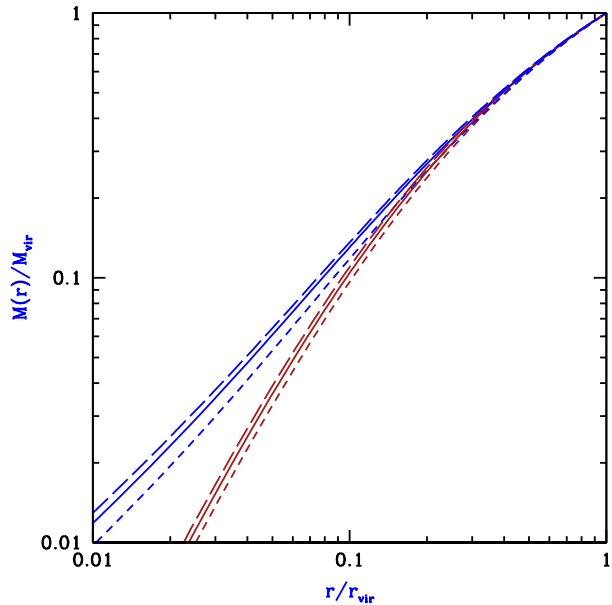}
\caption{Profiles of the enclosed 3-D mass in clusters as a function
of radius. We consider a host halo that follows an NFW model at
$z=0.3$, and assume satellites with $\cv=4$ at redshift $\zs=2$ that
make up a total fraction $\fs$ of the cluster's virial mass. We
consider a host $\cv=6.8$ and $\fs=20\%$ (short-dashed curves), a host
$\cv=7.5$ and $\fs=30\%$ (solid curves), or a host $\cv = 8.0$ and
$\fs=40\%$ (long-dashed curves). In each case we show the host profile
in the absence of cooling and adiabatic compression (bottom curve),
and the final profile (top curve) of a host that accreted satellites
that underwent gas cooling, adiabatic compression and stripping of
their dark matter halos before their baryons were redistributed
throughout the cluster. In each case $\fs$ is normalized so that the
host's dense core arises entirely from the satellite cores.}
\label{f:Mr}
\end{figure}

Figure~\ref{f:Mr} shows the effect of adiabatic compression on the 3-D
mass profiles of clusters. The effect is largest at the innermost
radii, where the satellites contribute a substantial fraction of the
halo mass. This is true even though our model in
equation~(\ref{e:final}) suppresses the density enhancement in this
$f_{\rm sat}(r_{\rm f}) \rightarrow 1$ limit; since the enclosed mass
changes rapidly with radius in the core, even a slight shift in the
stripping radius has a large effect on the mass profile. For the cases
considered, where the fractional mass contribution by satellites $\fs$
is 20, 30 or $40\%$, we find that the cluster profile at $r \la 0.1
\rv$ is substantially modified by adiabatic compression in the
satellites, even though the baryonic mass is assumed to have been
redistributed uniformly after the stripping. The enclosed mass at $r =
0.1 \rv$ is increased by $22\%$ (for $\fs=20\%$), $26\%$ (for
$\fs=30\%$), or $24\%$ (for $\fs=40\%$), and this enhancement factor
grows rapidly towards smaller radii. 

\begin{figure}%[th]
\includegraphics[width=84mm]{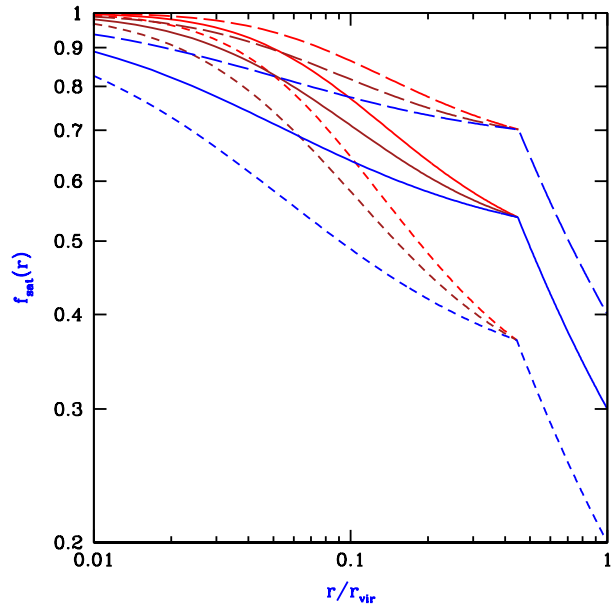}
\caption{Radial profile of the fractional satellite contribution to
the enclosed halo mass. Assumptions and notations are the same as in
Figure~\ref{f:Mr}, and in particular we consider $\fs=20\%$
(short-dashed curves), $30\%$ (solid curves), or $40\%$ (long-dashed
curves). In each case we show the satellite fraction in the absence of
cooling and adiabatic compression ($f_{\rm sat}^0(r)$, bottom curve),
the final fraction after stripping and baryon redistribution (based on
equation~(\ref{e:fsatf}), middle curve), and the higher fraction
present during stripping but still before baryon redistribution
(equation~(\ref{e:fsat}), top curve).}
\label{f:fsat}
\end{figure}

The model's features are illustrated in Figure~\ref{f:fsat}, which
shows the fractional satellite contribution to the halo mass profile.
In each case, the fraction is fixed to $\fs$ at the virial radius and
unity at $r \rightarrow 0$. There is a break in the curves at $r =
0.44 \rv$, which is the maximum cluster radius that can receive a
contribution from the satellite halos. This maximum arises from the
higher virial density at $z=2$ compared to the corresponding value at
$z=0.3$. At each radius, the satellite mass fraction reaches its
highest value during stripping, when it enjoys the enhancement due to
adiabatic compression but does not yet suffer the reduction due to the
redistribution of baryons. At $r = 0.1 \rv$ this fraction is between
60 and $90\%$, which makes our results at this radius (as seen in
Figure~\ref{f:Mr} and below) relatively insensitive to the precise
value of $\fs$; higher values of $\fs$ push the satellite fraction
higher towards unity, and in this limit stripping cannot significantly
increase the density of the satellite's dark matter, as reflected in
equation~(\ref{e:final}).

\begin{figure}%[th]
\includegraphics[width=84mm]{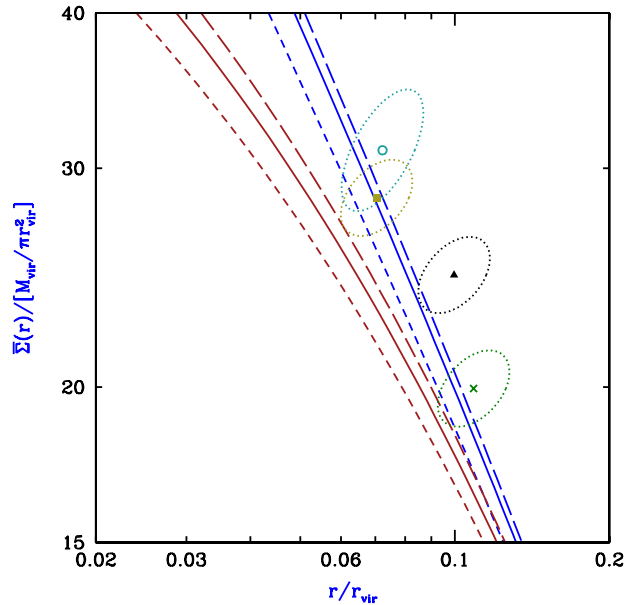}
\caption{Profile of the enclosed projected surface mass density versus
projected radius. Assumptions and notations are the same as in
Figure~\ref{f:Mr}, and in particular we consider $\fs=20\%$
(short-dashed curves), $30\%$ (solid curves), or $40\%$ (long-dashed
curves). For comparison, we show the observed values for four
clusters, A1689 (open circle), A1703 (square), Cl0024 (triangle), and
RXJ1347 ($\times$). For each cluster we show its critical density for
lensing versus effective Einstein radius, with the dot indicating a
central location and the ellipse showing the combined 1--$\sigma$
uncertainties due to the measurement errors in $\rE$ and $\Mv$.}
\label{f:Sigr}
\end{figure}

The implications of adiabatic compression for gravitational lensing
are displayed in Figure~\ref{f:Sigr}, which shows the profile of the
enclosed, projected 2-D mass density. We consider the same cluster and
satellite halo parameters as in the previous figures, and focus on the
range of projected radius corresponding to observed cluster Einstein
radii. For each of the observed clusters, we show the critical lensing
density versus effective Einstein radius, as a central point plus
1--$\sigma$ error ellipse. The figure is consistent with the factor of
$\sim 2$ discrepancy highlighted by \citet{BB08} between the observed
$\rE$ and the median theoretical prediction from pure dark matter
simulations; here the typical value of $r/\rv$ (for the bottom curves)
at the $\Sigma_{\rm cr}$ observed for each cluster is smaller only by
a factor of $\sim 1.5$, since our model has required us to adopt
somewhat higher than average cluster halo concentrations. Still, these
predictions in the absence of baryonic cooling for the most part lie
well outside the 1--$\sigma$ error ellipses of the observed clusters.

With the natural parameters that we have assumed for the satellites,
adiabatic compression resolves the current discrepancy; it boosts the
theoretical predictions enough to bring them well within the observed
error ellipses. For example, if $\fs=30\%$ then
for A1689, the observed (central) value $r/\rv = 0.072$ for the
effective Einstein radius can be compared with the predicted value (at
the same projected surface density equal to the critical lensing
density) of $r/\rv = 0.064$ (with adiabatic compression in the
satellites), and the previous pure dark-matter prediction (i.e.,
without adiabatic compression) of $r/\rv = 0.045$. For A1703,
the observed $r/\rv = 0.070$ can be compared with the theoretical
$r/\rv = 0.070$ (with adiabatic compression) and $r/\rv = 0.052$
(without). For Cl0024-17 the corresponding numbers are $r/\rv = 0.100$
compared to $r/\rv = 0.081$ (with) and $r/\rv = 0.064$ (without); for
RXJ1347, $r/\rv = 0.109$ compared to $r/\rv = 0.100$ (with) and $r/\rv
= 0.086$ (without). While the observed clusters still have slightly
high Einstein radii compared to the typical expected cluster profile,
the theoretical scatter in $\cv$ together with the observational
errors make the theoretical and observational predictions consistent
with each other.

As we have shown, our results depend only weakly on $\fs$, as long as
it is within a reasonable range. The results also depend slightly on
other assumed properties of the satellites. We illustrate this for A1689,
fixing $\fs=30\%$ and adjusting the host concentration accordingly in
each case. We find that lowering the satellite $\cv$ to 3 at $\zs=2$
decreases the predicted $r/\rv$ by $14\%$, while raising $\cv$ to 5
increases it by $12\%$. Assuming $\cv=3$ at $\zs=3$ raises the
predicted $r/\rv$ by $3\%$, while $\cv=5$ at $\zs=1$ lowers it by
$13\%$, all compared to our standard case of $\cv=4$ at $\zs=2$.
Finally, if we assume that only $50\%$ (rather than $100\%$) of the
baryons in the galactic satellites cooled and condensed before their
halos were stripped, i.e., in equation~(\ref{eq:Mf}) we use half
the cosmic fraction for $f_b$, then the predicted $r/\rv$ is reduced
by $9\%$ for A1689 and $\fs=30\%$. 

\section{Discussion}

We have demonstrated that dark matter compression due to baryonic
cooling inside galaxy halos can in turn lead to tidal stripping of
these galactic halos closer to the center of the galaxy cluster in
which they reside. Even if the baryons are later redistributed within
the cluster by feedback, a substantial effect remains due to the early
adiabatic compression. This effect can explain the high central mass
concentration of clusters in lensing observations. Our scenario, in
which only the inner $10$--$20\%$ of the virial radius is
significantly modified, is consistent with weak lensing measurements
at larger radii that find low cluster halo concentrations
\citep[e.g.,][]{Hirata}.

We have adopted a number of simplifying approximations in showing the
existence of the effect. Hydrodynamical simulations that avoid
overcooling of the baryons at the cluster core are necessary in order
to test our proposed mechanism in quantitative detail. Nevertheless,
our simplified treatment has demonstrated the general point that it is
possible for gas physics to significantly change the Einstein radius
of massive clusters, even without leaving a central baryon
concentration.

Finally, we note that ram pressure stripping of hot accreted baryons,
which make up the majority ($\sim 80\%$) of the cluster baryons, may
help to reduce the central baryon fraction. Indeed, cluster
simulations find a reduced baryon fraction at $r/\rv \la 0.2$
\citep[e.g.,][]{Kravtsov,Dolag} and match better the baryonic fraction
inferred from X-ray observations \citep{Vikhlinin}.

\section*{Acknowledgments}

RB is grateful for the kind hospitality of the {\it Institute for
Theory \& Computation (ITC)} at the Harvard-Smithsonian CfA, and also
acknowledges support by the Moore Distinguished Scholar program at
Caltech and the John Simon Guggenheim Memorial Foundation. This work
was supported in part by NASA grant NNX08AL43G, by Harvard University
funds and by BSF grant 2004386.

%\bsp

\label{lastpage}

\end{document}